\documentclass[journal]{IEEEtran}
\usepackage{amsfonts}
\usepackage{amssymb}
\usepackage{amsthm}
\usepackage{amsmath}
\usepackage{mathrsfs}
\usepackage{subfigure}
\usepackage{cite}
\usepackage{graphicx}
\usepackage{epstopdf}
\usepackage{enumerate}
\usepackage{color}
\usepackage{color, xcolor}
\usepackage{multirow}
\usepackage{booktabs}
\usepackage{algorithm}
\usepackage{algorithmic}

\renewcommand{\algorithmicrequire}{\textbf{Initialization:}}

\makeatletter

\newcommand\old[1]{}
\newtheorem{definition}{Definition}
\newtheorem{lemma}{Lemma}
\newtheorem{corollary}{Corollary}

\newtheorem{theorem}{Theorem}
\newtheorem{remark}{Remark}
\newtheorem{example}{Example}
\newtheorem{assumption}{Assumption}

\makeatother

\begin{document}

\date{\today}

\title{Fast Subspace Identification Method Based on Containerised Cloud Workflow Processing System}
\author{Runze Gao, Yuanqing Xia$^{*}$, Guan Wang, Liwen Yang, Yufeng Zhan
\thanks{R. Gao, Y. Xia, G. Wang, L. Yang and Y. Zhan are with the School of Automation, Beijing Institute of Technology, Beijing 100081, China. G. Wang is also with the School of Information Science and Engineering, University of Zaozhuang, Shandong 277160, China. ({\footnotesize {\em Corresponding
author: Yuanqing Xia}). Email address:
gaorunze$\_$bit@163.com (R. Gao), xia$\_$yuanqing@bit.edu.cn (Y. Xia), netspecters@126.com (G. Wang), ytyangliwen@163.com (L. Yang), yu-feng.zhan@bit.edu.cn (Y. Zhan)
}}}
\markboth{manuscript for review}{}
\maketitle

\begin{abstract}
Subspace identification (SID) has been widely used in system identification and control fields since it can estimate system models only relying on the input and output data by reliable numerical operations such as singular value decomposition (SVD). However, high-dimension Hankel matrices are involved to store these data and used to obtain the system models, which increases the computation amount of SID and leads SID not suitable for the large-scale or real-time identification tasks. In this paper, a novel fast SID method based on cloud workflow processing and container technology is proposed to accelerate the traditional algorithm. First, a workflow-based structure of SID is designed to match the distributed cloud environment, based on the computational feature of each calculation stage. Second, a containerised cloud workflow processing system is established to execute the logic- and data- dependent SID workflow mission based on Kubernetes system. Finally, the experiments show that the computation time is reduced by at most $91.6\%$ for large-scale SID mission and decreased to within 20 ms for the real-time mission parameter.

\end{abstract}


\section{Introduction}\label{Introduction}

\IEEEPARstart{I}{n} the decades, subspace identification (SID) methods have attracted substantial interest in system identification and control fields. Several important algorithms including CVA, N4SID, MOSEP and data-driven predictive control have been developed \cite{van2012subspace, van1994n4sid, xia2013data, yu2019subspace}. SID methods enable accurate identification of state-space models using the reliable matrix operations such as singular value decomposition (SVD), and the state-space models can be consistently identified from the input and output data under mild conditions. In SID methods, high-dimension and low-rank Hankel matrices are derived to store the historical data and then obtain the final system model. Its major drawback, on the other hand, is that the involved large-scale data matrices increase the computation amount of matrix operations and this leads SID methods not suitable for the large-scale or real-time identification tasks.

There have been several methods to reduce the computation time of SID methods. The majority of them focus on the improvement of the numerical algorithm design. For example, recursive identification (RSI) has been proposed to update the observability matrix estimation using new data \cite{oku2002recursive, houtzager2011recursive}. An accelerated stochastic SID method has been presented to compute the model of the high-order systems with noise \cite{dohler2013efficient}. The minority study how to improve the computation effectiveness via more powerful computing devices. A novel method to speedup the large-scale ambient oscillation identification using multi-cores server has been addressed in \cite{wu2016fast}. This kind of methods have advantages over the numerical improvements on utilizing the computing power of multi-processers structure. But \cite{wu2016fast} is still restricted to the traditional local devices of which the computing resources such as CPU, memory and etc are fixed and lack of elasticity. 

\begin{figure}[!ht]
  \centering
  \includegraphics[height=1.3in]{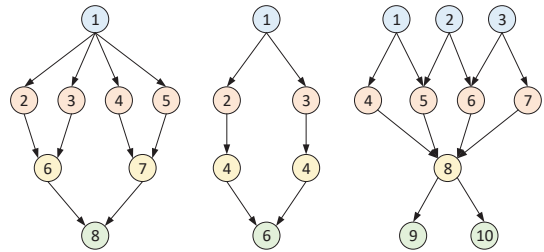}\\
  \caption{Examples of the Cloud Workflow Structure}\label{Examples of the Cloud Workflow Structure}
  \vspace{-0.9em}
\end{figure}

Nowadays, many scientific missions such as deep learning, genetic calculation and earthquake wave analysis have been processed in cloud server because of the computing ability of cloud computing \cite{senyo2018cloud, chen2016parallel, xiao2020malfcs, xia2020cloud}. Subspace identification is also a kind of scientific mission of which the calculation is highly time-consuming. However, there not exists the combined work of cloud computing and the SID mission. Furthermore, the computing mode of the SID mission is centralized but the structure of cloud environment is distributed. If the SID mission is deployed in cloud server directly, the computation efficiency would not be improved since the computing ability of distributed structure is not utilized. Therefore, this paper proposes the workflow-based SID method, which restructure the SID mission into the cloud worklfow form for matching the distributed cloud environment. As shown in Fig. \ref{Examples of the Cloud Workflow Structure}, the mentioned cloud worklfows are represented graphically by directed acyclic graphs (DAGs). The nodes represent computational tasks and the directed edges between the nodes determine the inter-dependencies between the tasks \cite{wu2015workflow, zhang2017bi, zhu2018scheduling}. In cloud environment, the tasks in a workflow would be scheduled to different processing nodes of which the required computing resources are encapsulated from a shared resource pool.

Therefore, a fast, stable and flexible scheme of encapsulating computing resources is required to execute workflow tasks. Cloud workflow processing based on virtual machine (VM) is the present major approach \cite{haidri2020cost, zhu2018scheduling, mao2019learning}. VM technology provides a virtual operation system to access computing resources and create isolated spaces for executing the tasks. However, the operation overheads of VM is quite high, which contains the creation time, startup time and configuring time as well as complexity. For example, the creation and startup stages of VM cost several minutes and 30-40 s, respectively. To reduce the overheads, container technology is created and applied \cite{kaur2017container, ranjan2020energy, goldschmidt2018container, mellado2020container}. This technology enables that the smaller isolated processes for processing tasks can be created as well as started in an instant, and released automatically when the tasks are finished. Compared with VM technology, container also has potential benefits on the configuring complexity, computing resource usage, overhead cost, migration speed across hosts and template sharable ability as summarized in Table \ref{Benefits of Container Technology} \cite{ranjan2020energy}.

\begin{table}[htbp]
	\centering
	\caption{Benefits of Container Technology}
	\begin{tabular}{|l|c|c|}
		\toprule  
		\textbf{Parameter}&\textbf{Container technology}&\textbf{VM technology} \\
        \hline
        \hline
		Kernel state&Booting needed&Operational \\
        \hline
        Creation time&$<$ 1 s&Minutes \\
        \hline
        Start/stop time&$<$ 50 ms&30-40 and 5-10 s \\
        \hline
        Startup overhead&Low&Relatively high \\
        \hline
        Configuring complexity&Midden&High \\
        \hline
        CPU/memory usage&Low&High \\
        \hline
        Migration&Quick&Slow \\
        \hline
        Template sharing&Yes&No \\
		\bottomrule  
	\end{tabular}
\label{Benefits of Container Technology}
\end{table}

Motivated by this, to improve the computation efficiency of SID methods, by adopting cloud workflow processing approach and container technology, the main contributions of this paper on the novel method are summarized in the following.
\begin{enumerate}
  \item A workflow-based structure of the SID method is established to match the distributed cloud environment, based on the computational feature of each calculation stage.
  \item A containerised cloud workflow processing system is established to execute the logic- and data- dependent SID workflow missions based on Kubernetes system.
  \item Based on this system, the Monte Carlo experiments are carried out and the results show that the computation time is reduced by at most $91.6\%$ for large-scale SID mission and decreased to within 20 ms for the real-time mission parameter.
\end{enumerate}

The rest of this paper is organized as follow. Section \ref{Preliminary} presents the preliminary of SID. Section \ref{Overview of cloud-based subspace identification section} provides the overview of the cloud-based SID method. The computational features analyses and the establishment of the SID workflow are presented in Section \ref{Fast subspace identification method}. The design and establishment of the containerised cloud workflow processing are provided in Section \ref{Containerised cloud workflow processing system}. The experiment evaluation as well as the result analyses are provided in Section \ref{Performance Evaluation}. The final section presents the conclusion and future work.

\section{Preliminary of Subspace Identification}\label{Preliminary}

\begin{algorithm}[th] 
    \renewcommand{\algorithmicrequire}{\textbf{Input:}}
	\renewcommand{\algorithmicensure}{\textbf{Output:}}
    \caption{\textbf{N4SID: A Deterministic Subspace System Identification Method}}
    \label{N4SID: A Deterministic Subspace System Identification Method}
    \small{\textbf{Input:}
    The scale parameters $N,j$ and the Hankel matrices of inputs and outputs $U_p, U_f, U_{p}^{+}, U_{f}^{-}, Y_p, Y_f, Y_{p}^{+}, Y_{f}^{-}$.\\
    \textbf{Output:} The identified state-space matrices: $A$, $B$, $C$ and $D$.}
    \begin{algorithmic}[1]
    \small{
        \STATE Calculate the oblique projections:
        \begin{eqnarray*}
          \small{ \mathcal{O}_{i}} &=& \small{Y_{f}/_{U_f}\textbf{W}_{p}}, \\
          \small{\mathcal{O}_{i-1}} &=& \small{Y_{f}^{-}/_{U_{f}^{-}}\textbf{W}_{p}^{+}}.
        \end{eqnarray*}
        \STATE Calculate the SVD of the weighted oblique projection:
        \begin{equation*}
          \small{ W_{1}\mathcal{O}_{i}W_{2} = USV^{T}.}
        \end{equation*}
        \STATE Determine the order by inspecting the singular values of $S$ and partition the SVD accordingly to obtain $U_{1}$, $S_{1}$.
        \begin{equation*}
          \small{ USV^{T} = \left[
                      \begin{array}{cc}
                        U_1 & U_2 \\
                      \end{array}
                    \right]\left[
                             \begin{array}{cc}
                               S_1 & 0 \\
                               0 & S_2 \\
                             \end{array}
                           \right]\left[
                                    \begin{array}{c}
                                      V_{1}^{T} \\
                                      V_{2}^{T} \\
                                    \end{array}
                                  \right].}
        \end{equation*}
        \STATE Determine the extended matrices $\Gamma^{\dag}_{i}$ and $\Gamma^{\dag}_{i-1}$ and estimated state sequences $X_{i}$ and $X_{i+1}$.
        \STATE Solve the set of linear equations for $A$, $B$, $C$ and $D$:
        \begin{equation*}
          \small{\left[
            \begin{array}{c}
              X_{i+1} \\
              Y_{i} \\
            \end{array}
          \right]=\left[
                    \begin{array}{cc}
                      A & B \\
                      C & D \\
                    \end{array}
                  \right]\left[
                           \begin{array}{c}
                             X_{i} \\
                             U_{i} \\
                           \end{array}
                         \right].}
        \end{equation*}}
    \end{algorithmic}
\end{algorithm}

In this preliminary, the standard SID algorithm is recalled. Consider the following deterministic model to be identified:
\begin{eqnarray}\label{system}
  x(k+1) &=& Ax(k)+Bu(k) \\
  y(k) &=& Cx(k)+Du(k)
\end{eqnarray}
where the system matrices $A\in {\bf R}^{n\times n}$, $B\in {\bf R}^{n\times m}$, $C\in {\bf R}^{l\times n}$, $D\in {\bf R}^{l\times m}$.

In the non-sequential data processing, Hankel matrices are always used to store and deal with data. For example, the Hankel matrices of the input data series $\{u(0),u(1),\ldots,u(2N+j-2)\}$ are defined as
\begin{eqnarray}
\centering
  \small{\!\!\!\!\!\!U_p \!\!\!\!}&=& \small{\!\!\!\!\!\!\left[
            \begin{array}{cccc}
              \!\!u(0) & \!\!u(1) & \!\!\ldots & \!\!u(j-1) \\
              \!\!u(1) & \!\!u(2) & \!\!\ldots & \!\!u(j) \\
              \!\!\vdots & \!\!\vdots & \!\!\ddots & \!\!\vdots \\
              \!\!u(N-1) & \!\!u(N) & \!\!\ldots & \!\!u(N+j-2) \\
            \end{array}
          \right]}, \\
  \small{\!\!U_f \!\!\!\!}&=& \small{\!\!\!\!\!\!\left[
            \begin{array}{cccc}
              \!\!\!u(N) & \!\!\!u(N+1) & \!\!\!\ldots & \!\!\!u(N+j-1) \\
              \!\!\!u(N+1) & \!\!\!u(N+2) & \!\!\!\ldots & \!\!\!u(N+j) \\
              \!\!\!\vdots & \!\!\!\vdots & \!\!\!\ddots & \!\!\!\vdots \\
              \!\!\!u(2N-1) & \!\!\!u(2N) & \!\!\!\ldots & \!\!\!u(2N+j-2) \\
            \end{array}
          \right]}
\end{eqnarray}
where the subscript $``\emph{p}"$ stands for $``$past$"$, $``\emph{f}"$ means $``$future$"$ and $N,$ $j\in \mathbb{N}^{+}$. The Hankel matrices of $\{y(0),y(1),\ldots,y(2N+j-2)\}$ are written likewise, denoted as $Y_p$ and $Y_f$. The matrix containing the $U_p$ and $Y_p$ is $\textbf{W}_p$:
\begin{equation}
  \small{\textbf{W}_p = \left[
          \begin{array}{c}
            Y_p \\
            U_p \\
          \end{array}
        \right]}
\end{equation}

\begin{figure*}[!ht]
  \centering
  \includegraphics[height=3.3in]{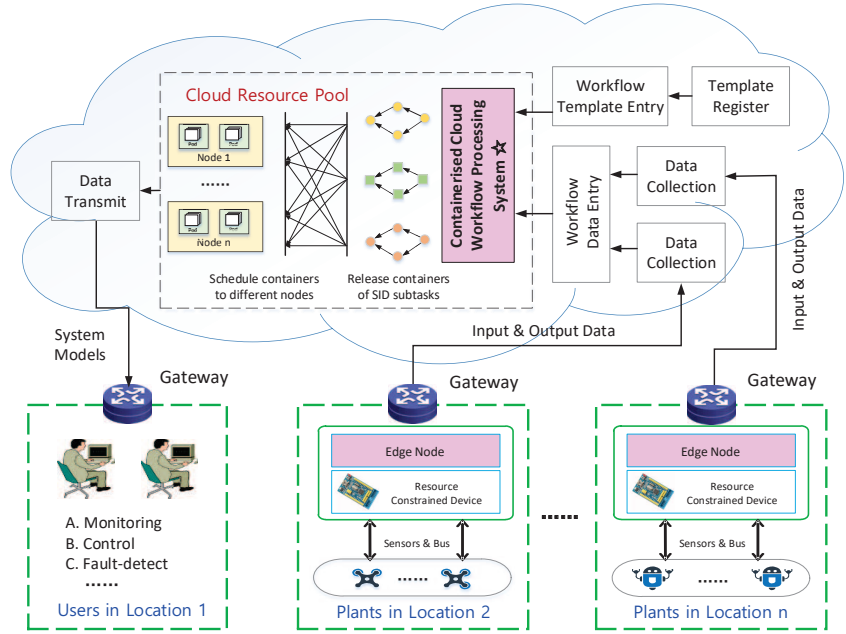}\\
  \caption{The Overview of Cloud-based Subspace Identification}\label{Overview of cloud-based subspace identification}
\end{figure*}

To keep the convergence of the identified results, the parameter $j$ is required to be much larger than $N$ \cite{van2012subspace}. Thus the Hankel matrices in system identification are always high-dimension and low-rank, which inspires the authors to further handle the matrices. As one of the classical SID methods, N4SID is summarized as Algorithm 1 \cite{van1994n4sid}, where the oblique projection and other new variables are defined in Appendix \ref{Definitions in N4SID Algorithm} and the details of the step 4 are described in Appendix \ref{Step 4 in N4SID Algorithm}.

\section{Overview of cloud-based subspace identification}\label{Overview of cloud-based subspace identification section}

In this section, the overview of the cloud-based subspace identification is provided, as shown in Fig. \ref{Overview of cloud-based subspace identification}. This overview consists of three layers, which are cloud platform, terminal and users layers. Some important modules are described in Table \ref{Descriptions of part defined modules}. Cloud platform layer is the upper part of this overview. After receiving the input and output data from the edge nodes, the data collection modules transmit these data to the wrokflow data entry. In the meanwhile, the SID workflow template is pulled to the workflow template entry from the template register. Then the data and workflow structure as well as task images are received by the containerised cloud workflow processing system (with the star symbol in Fig. \ref{Overview of cloud-based subspace identification}) which is designed in Section \ref{Containerised cloud workflow processing system}. This system creates containers and schedules them to different computing nodes to keep load balance. Finally, the identified models are obtained and transmitted to the users node.

Terminal layer is in the lower right of this overview. There would be multiple edge nodes distributed in different geographic positions. Each edge node is a relatively independent system closed to the terminal plants which generate original data. However, the computing ability is constrained in the edge nodes by the devices cost, deployment space and energy supply. Thus it is beneficial to bring the computation-intensive SID missions to the cloud platform. The users layer is in the upper left of this overview. In the users node, the received models can be applied in various functions such as system monitoring, data-driven control, fault-detecting and etc.

\begin{table}[htbp]
	\centering
	\caption{Descriptions of part defined modules in Fig. \ref{Overview of cloud-based subspace identification}}
    \renewcommand{\multirowsetup}{\centering}
	\begin{tabular}{|c|c|}
		\toprule  
		\textbf{Module}&\textbf{Description} \\
		\hline  
        \hline
        \multirow{2}{1.7cm}{Cloud Resource Pool}&\multirow{2}{6.4cm}{A abstract concept for describing all computing resource including physical server, VM and container}\\
        & \\
        \hline
        \multirow{2}{1.7cm}{Template Register}&\multirow{2}{6.4cm}{A sharing warehouse for storing reusable images of pre-defined tasks in SID method}\\
        & \\
        \hline
        \multirow{2}{1.7cm}{Workflow Template}&\multirow{2}{6.4cm}{The template consisting of the task images and the topology structure between the tasks}\\
        & \\
        \hline
        \multirow{2}{1.8cm}{Workflow Template Entry}&\multirow{2}{6.4cm}{A middle module to provide the template of the SID workflow pulled from Docker Register}\\
        & \\
        \hline
        \multirow{2}{1.7cm}{Workflow Data Entry}&\multirow{2}{6.4cm}{A middle module to provide the input and output data of the identified plant for the SID workflow}\\
        & \\
		\bottomrule  
	\end{tabular}
\label{Descriptions of part defined modules}
\end{table}

\begin{table*}[htbp]
	\centering
	\caption{Averaged calculation time statistics for topical parameters in Monte Carlo experiments}
    \renewcommand{\multirowsetup}{\centering}
	\begin{tabular}{|c|c|c|c|c|c|c|c|c|c|c|c|}
		\hline  
        \cline{1-10}
        \multirow{4}{1.5cm}{\textbf{Data scale}}&\multirow{4}{2cm}{\textbf{Parameters}}&\multicolumn{2}{c|}{\multirow{2}{*}{\textbf{Oblique projection}}}&  \multicolumn{2}{c|}{\multirow{2}{2.2cm}{\textbf{SVD}}}  & \multicolumn{2}{c|}{\multirow{2}{2.2cm}{\textbf{Estimation}}} &\multirow{4}{1.1cm}{\textbf{Total (s)}} & \multirow{4}{1.9cm}{\textbf{Major stage}}  \cr
        &&  \multicolumn{2}{c|}{\multirow{2}{2.2cm}{}}  & \multicolumn{2}{c|}{\multirow{2}{2.2cm}{}} &\multicolumn{2}{c|}{\multirow{2}{2.2cm}{}} & &\cr
        \cline{3-8}
        &&\multirow{2}{1.1cm}{Time (s)}&\multirow{2}{1.1cm}{Per.}&\multirow{2}{1.1cm}{Time (s)}&\multirow{2}{1.1cm}{Per.}&\multirow{2}{1.1cm}{Time (s)}&\multirow{2}{1.1cm}{Per.}&&\\
        &&&&&&&&& \\
        \cline{1-10}
        \cline{1-10}
        \cline{1-10}
		\multirow{2}{1.5cm}{Small-scale}
        &$N=10,j=300$&0.0090&\textbf{67.7$\%$}&0.0040&30.1$\%$&0.0003&2.3$\%$&0.0133&\multirow{2}{*}{Oblique projection} \\
        &$N=20,j=300$&0.0315&\textbf{77.0$\%$}&0.0092&22.5$\%$&0.0002&0.5$\%$&0.0409&\multirow{2}{1.9cm}{}\\
        \hline
        \hline
        \multirow{2}{*}{Middle-scale}
        &$N=10,j=1000$&0.0106&20.5$\%$&0.0408&\textbf{78.8$\%$}&0.0004&0.8$\%$&0.0518&\multirow{2}{1.9cm}{SVD} \\
        &$N=20,j=1000$&0.0348&33.9$\%$&0.0676&\textbf{65.8$\%$}&0.0004&0.0$\%$&0.1028&\multirow{2}{1.9cm}{}\\
        \hline
        \hline
        \multirow{4}{1.5cm}{Large-scale}
        &$N=10,j=10000$&0.0316&0.7$\%$&4.5347&\textbf{99.3$\%$}&0.0009&0.0$\%$&4.5672&\multirow{4}{1.9cm}{SVD}\\
        &$N=20,j=10000$&0.0810&1.1$\%$&6.9837&\textbf{98.8$\%$}&0.0019&0.1$\%$&7.0666&\multirow{2}{1.9cm}{}\\
        \cline{2-9}
        \multirow{4}{1.5cm}{}
        &$N=50,j=10000$&0.2666&2.0$\%$&13.1051&\textbf{98.0$\%$}&0.0024&0.0$\%$&13.3741&\multirow{2}{1.9cm}{}\\
        &$N=50,j=20000$&0.4375&0.7$\%$&66.2683&\textbf{99.3$\%$}&0.0531&0.0$\%$&66.7589&\multirow{2}{1.9cm}{}\\
        \hline
        \cline{1-10}
	\end{tabular}
\label{Mean calculation time statistics for topical parameters in Monte Carlo experiments}
\end{table*}


\section{Fast workflow-based subspace identification method}\label{Fast subspace identification method}

In this section, the computational features of all calculation stages in Algorithm \ref{N4SID: A Deterministic Subspace System Identification Method} are analysed in detail. The results show that SVD is the most time-consuming stage. Then, we carry out the parallelization of subspace identification method based on the truncated SVD algorithm. Later, the workflow of subspace identification method in DAG form is established. Finally, the computational complexity analysis is provided.

\subsection{Feature analysis of each calculation stage}

To discuss the computational features, Algorithm \ref{N4SID: A Deterministic Subspace System Identification Method} is divided into three stages, which are
\begin{itemize}
  \item Oblique projection: compute the oblique projections $\mathcal{O}_{i},\mathcal{O}_{i-1}$.
  \item SVD: compute the SVD and obtain the system order $n$ and decomposed results $U_1,S_1$.
  \item Estimation: compute the extended matrices $\Gamma^{\dag}_{i},\Gamma^{\dag}_{i-1}$, the estimated state sequences $X_{i},X_{i+1}$ and estimate the system matrices $A,B,C,D$.
\end{itemize}

Further, to analyse the averaged calculation time of each stage, a group of Monte Carlo experiments with 100 times repetition of a two-order system are carried out in a local computer with i7-3770 CPU and 8 GB memory. In Table \ref{Mean calculation time statistics for topical parameters in Monte Carlo experiments}, the statistical results are list by dividing into three stages, i.e., the small-, middle- and large- scales. In the small-scale stage, the stage of oblique projection is the main time-consuming process. But the total calculation time is relatively less, so it's unnecessary to construct workflow for this stage.

With the data scale increasing, the cost time of SID method grows and becomes hard to ignore. In the middle-scale stage, of which the parameters are always used in SID-based data-driven predictive control, SVD occupies the major stage as the percentage is over 65$\%$. In the large-scale stage, the percentage of SVD is over 98$\%$. That is to say, when the data amount reaches the middle- and large- scales, SVD is the main time-consuming stage and the total calculation times are relatively high. Thus the speedup is indeed desired.

\subsection{Parallelization based on the truncated SVD algorithm}

To accelerate the computation of SID method, we consider it in a distributed parallelization framework. Based on the analysis in the last subsection, the SVD stage would be reconstructed into a DAG, as the main part of the whole SID workflow. The truncated SVD algorithm can achieve the distributed decomposition in a lightweight way \cite{bjorck2015numerical}, which inspires us to establish the DAG of SVD.

Assume an $m\times n$ matrix $A$ is split column-wise into submatrices $A_1$ and $A_2$ with the sizes $m\times n_1$ and $m\times n_2$, respectively and $n_1+n_2 =n$. Conduct SVD on $A_1$, $A_2$ and obtain the results $A_1=U_1\Sigma_1V_1^{T}$, $A_2=U_2\Sigma_2V_2^{T}$. Then the SVD of $A = [A_1\ A_2]$ can be written as\vspace{-0.1em}
\begin{equation}
  \small{\left[\!
    \begin{array}{cc}
      A_1\! & \!A_2 \\
    \end{array}
  \!\right]
   \!=\! \left[\!
    \begin{array}{cc}
      U_1\Sigma_1\! & \!U_2\Sigma_2 \\
    \end{array}
  \!\right]\left[\!
                                             \begin{array}{cc}
                                               V_1^{T}\! & \!\textbf{0}\! \\
                                               \textbf{0}\! & \!V_2^{T}\! \\
                                             \end{array}
                                           \!\right]
   \\
   \!\!=\! E \left[\!
                                             \begin{array}{cc}
                                               V_1^{T}\! & \!\textbf{0}\! \\
                                               \textbf{0}\! & \!V_2^{T}\! \\
                                             \end{array}
                                           \!\right]}\vspace{-0.1em}
\end{equation}
where $E = \left[\!
    \begin{array}{cc}
      U_1\Sigma_1\!\! & \!U_2\Sigma_2 \\
    \end{array}
  \!\right]$. Conduct SVD on $E$ and obtain\vspace{-0.1em}
\begin{equation}
  \small{\left[\!
    \begin{array}{cc}
      A_1\! & \!A_2 \\
    \end{array}
  \!\right] \!=\! U\Sigma \widetilde{V}^{T} \; \left[\!
                                             \begin{array}{cc}
                                               V_1^{T}\! & \!\textbf{0}\! \\
                                               \textbf{0}\! & \!V_2^{T}\! \\
                                             \end{array}
                                           \!\right] \\
   \!=\! U\Sigma V^{T}}\vspace{-0.1em}
\end{equation}
where $V^{T} = \widetilde{V}^{T}\left[\!
                                             \begin{array}{cc}
                                               V_1^{T}\! & \!\textbf{0}\! \\
                                               \textbf{0}\! & \!V_2^{T}\! \\
                                             \end{array}
                                           \!\right]$ is the product of two orthogonal matrices and hence is also an orthogonal matrix.
\begin{algorithm}[th]
    \renewcommand{\algorithmicrequire}{\textbf{Input:}}
	\renewcommand{\algorithmicensure}{\textbf{Output:}}
    \caption{\textbf{Distributed Truncated SVD Algorithm}}
    \label{Distributed truncated SVD algorithm}
    \small{\textbf{Input:} A matrix to be decomposed $A_{m\times n}$, block width $col$.\\
    \textbf{Output:} Decomposed results $U$, $S$ and $V$.}

    \begin{algorithmic}[1]
    \small{
         \STATE \textbf{\textrm{function}} \textsc{ParallelSVDbyCols}($A_{m\times n}$, $col$):
         \STATE \qquad $N_c$ = round($n / col$ + 0.45)
         \STATE \qquad $l_A$ = list(); $l_U$ = list(); $l_\Sigma$ = list(); $l_V$ = list()
         \STATE \qquad $l_A$ is filled with $N_c$ column blocks of $A$
         \STATE \qquad \textbf{\textrm{for}} $i$ in range($N_c$) \textbf{\textrm{do}}
         \STATE \qquad \qquad $U_i$, $\Sigma_i$, $V_i$ = SVD$(l_A(i))$
         \STATE \qquad \qquad $l_U$ += $U_i$; $l_\Sigma$ += $\Sigma_i$; $l_V$ += $V_i$
         \STATE \qquad \textbf{\textrm{end for}}
         \STATE \qquad return $\hat{U}$, $\hat{\Sigma}$, $\hat{V}$ = DoMergeOfBlocks($l_U$, $l_\Sigma$, $l_V$)
         \STATE \textbf{\textrm{end function}}
         \STATE \textbf{\textrm{function}} \textsc{DoMergeOfBlocks}($l_U$, $l_\Sigma$, $l_V$):
         \STATE \qquad $Nl$ = len($l_U$)
         \STATE \qquad $level$ = ceil($\log_{2}Nl$)
         \STATE \qquad \textbf{\textrm{for}} $i$ in range($level$) \textbf{\textrm{do}}
         \STATE \qquad \qquad $l_{Ut} = l_{U}$; $l_{\Sigma t} = l_{\Sigma}$; $l_{Vt} = l_{V}$
         \STATE \qquad \qquad $l_U$ = list(); $l_\Sigma$ = list(); $l_V$ = list()
         \STATE \qquad \qquad \textbf{\textrm{for}} $j$ in range($0$, $Nl$, $2$) \textbf{\textrm{do}}
         \STATE \qquad \qquad \qquad $U_j$, $\Sigma_j$, $V_j$ = BlockMerge($l_{Ut}(j)$, $l_{\Sigma t}(j)$,\\ \qquad \qquad \qquad \quad $l_{Vt}(j)$, $l_{Ut}(j+1)$, $l_{\Sigma t}(j+1)$, $l_{Vt}(j+1))$
         \STATE \qquad \qquad \qquad $l_U$ += $U_j$; $l_\Sigma$ += $\Sigma_j$; $l_V$ += $V_j$
         \STATE \qquad \qquad \textbf{\textrm{end for}}
         \STATE \qquad \qquad \textbf{\textrm{if}} $Nl$ is odd \textbf{\textrm{then}}
         \STATE \qquad \qquad \qquad Append the last elements of $l_{Ut}$, $l_{\Sigma t}$ \\ \qquad \qquad \qquad \quad and $l_{Vt}$ to $l_{U}$, $l_{\Sigma}$ and $l_{V}$, respectively
         \STATE \qquad \qquad \textbf{\textrm{end if}}
         \STATE \qquad \textbf{\textrm{end for}}
         \STATE \qquad return $\hat{U}$, $\hat{\Sigma}$, $\hat{V}$
         \STATE \textbf{\textrm{end function}}
         \STATE \textbf{\textrm{function}} \textsc{BlockMerge}($U_1$, $\Sigma_1$, $V_1$, $U_2$, $\Sigma_2$, $V_2$):
         \STATE \qquad $U_{1_{k}}$, $\Sigma_{1_{k}}$, $V_{1_{k}}$ = DoTruncate($U_1$, $\Sigma_1$, $V_1$)
         \STATE \qquad $U_{2_{l}}$, $\Sigma_{2_{l}}$, $V_{2_{l}}$ = DoTruncate($U_2$, $\Sigma_2$, $V_2$)
         \STATE \qquad $U_{r}$, $\Sigma_{r}$, $\hat{V}_{r}$ = SVD($[U_{1_{k}}\Sigma_{1_{k}}\ U_{2_{l}}\Sigma_{2_{l}}]$)
         \STATE \qquad $V_{r}$ = $\hat{V}*blkdiag(V_{1_{r}}, V_{2_{l}})$
         \STATE \qquad return $U_{r}$, $\Sigma_{r}$, $V_{r}$
         \STATE \textbf{\textrm{end function}}
         \STATE \textbf{\textrm{function}} \textsc{DoTruncate}($U$, $\Sigma$, $V$):
         \STATE \qquad $k$ = $rank$($\Sigma$); $U_{k}$ = $U$(:,1:$k$)
         \STATE \qquad $\Sigma_{k}$ = $\Sigma$(1:$k$,1:$k$); $V_{k}$ = $V$(1:$k$,1:$k$)
         \STATE \qquad return $U_{k}$, $\Sigma_{k}$, $V_{k}$
         \STATE \textbf{\textrm{end function}}}
    \end{algorithmic}
    \vspace{-0.25em}
\end{algorithm}

In SID method, the matrices to be decomposed by singular values are high-dimensional since these matrices are derived and combined by the original Hankel matrices. But since $j$ is required to be much larger than $N$, the row lengths of these matrices are much less than the column widths. That is to say, these matrices are low-rank matrices in the meanwhile. Therefore, the low rank approximation method can be used here to discard the low-value data in each level of decomposition, which is beneficial to reduce the computation.
\begin{figure}[!ht]
  \centering
  \includegraphics[width=2.5in]{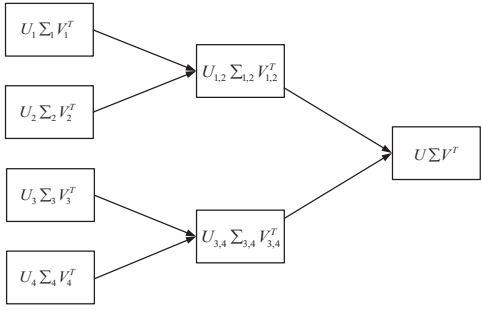}\\
  \caption{The Column-based Distributed SVD Method by MAT Operation}\label{SVD}
\end{figure}
\begin{figure*}[!ht]
  \centering
  \includegraphics[width=5in]{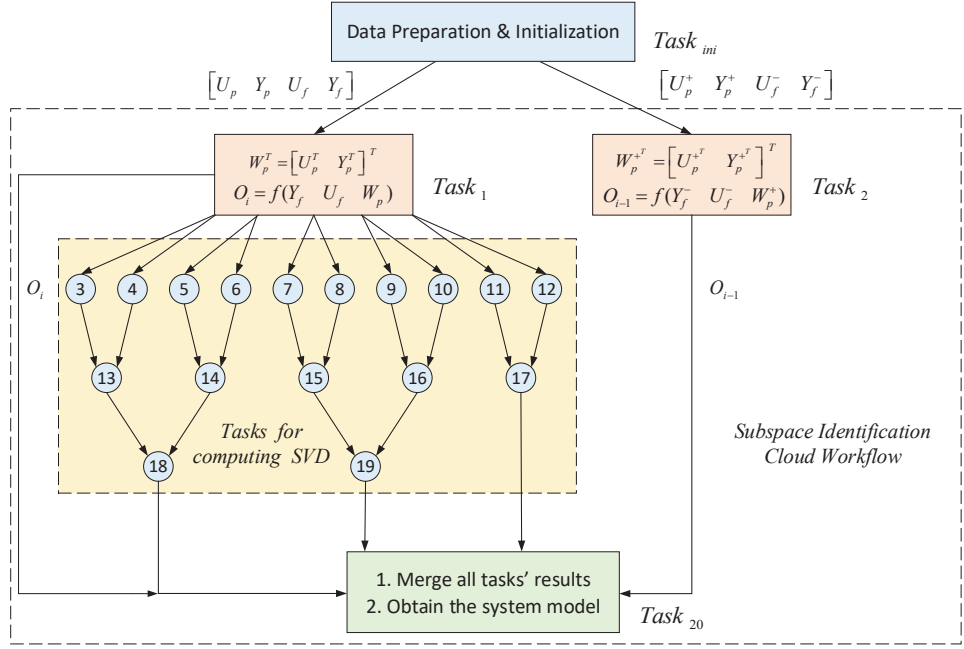}\\
  \caption{The Workflow Structure of Subspace Identification Method ($MPT = 10$)}\label{Workflow Structure of Subspace Identification Method}
\end{figure*}
The low rank approximations would be conducted after the individual SVDs are finished, of which the criteria is that only the square matrices corresponding to the singular values are retained. Assume $rank(A)=r$, $rank(A_1)=k$ and $rank(A_2)=l$. Thus, $A_1\!\approx\! U_{1_{k}}\Sigma_{1_{k}}V_{1_{k}}^{T}$ and $A_2\!\approx \!U_{2_{l}}\Sigma_{2_{l}}V_{2_{l}}^{T}$ indicate rank-$k$ and rank-$l$ approximations, respectively. The result is truncated to a rank-$r$ approximation as follows.\vspace{-0.35em}
\begin{equation}
  \small{\left[\!
    \begin{array}{cc}
      A_1\! & \!A_2 \\
    \end{array}
  \!\right] \!\approx \! \left[\!
    \begin{array}{cc}
      U_{1_{k}}\Sigma_{1_{k}}\! & \!U_{2_{l}}\Sigma_{2_{l}} \\
    \end{array}
  \!\right]\!\left[\!
                                             \begin{array}{cc}
                                               V_{1_{k}}^{T}\! & \!\textbf{0}\! \\
                                               \textbf{0}\! & \!V_{2_{l}}^{T}\! \\
                                             \end{array}
                                           \!\right]\! \\
   \!= \!U_r\Sigma_rV_r^{T}}.\vspace{-0.35em}
\end{equation}

\begin{table}[htbp]
	\centering
	\caption{Relationships between Tasks in Workflow and Functions}
    \renewcommand{\multirowsetup}{\centering}
	\begin{tabular}{|c|c|c|c|}
		\toprule  
		\textbf{Image}&\textbf{Index}&\textbf{Level}&\textbf{Function} \\
        \hline
        \hline
		\multirow{2}{1.15cm}{$Image_{ini}$}&\multirow{2}{1.2cm}{$Task_{ini}$}&\multirow{2}{0.45cm}{$-$} &\multirow{2}{4.2cm}{Prepare Hankel matrices by creating history data}\\
        &&& \\
        \hline
        \multirow{2}{1cm}{$Image_A$}&\multirow{2}{1.2cm}{$Task_1$}&\multirow{2}{0.45cm}{1}&\multirow{2}{4.2cm}{Generate $O_i$ for SVD tasks} \\
        &&& \\
        \hline
        \multirow{2}{1cm}{$Image_B$}&\multirow{2}{1.2cm}{$Task_2$}&\multirow{2}{0.45cm}{1}&\multirow{2}{4.2cm}{Generate $O_{i-1}$ for calculating extended state matrix $X_{i+1}$} \\&&& \\
        \hline
        \multirow{2}{1cm}{$Image_C$}&\multirow{2}{1.2cm}{$Task_3-Task_{12}$}&\multirow{2}{0.45cm}{2} &\multirow{2}{4.2cm}{Compute the truncated SVD of \\each block matrices of $O_i$}\\&&& \\
        \hline
        \multirow{2}{1cm}{$Image_D$}&\multirow{2}{1.2cm}{$Task_{13}-Task_{19}$}&\multirow{2}{0.45cm}{3, 4}&\multirow{2}{4.2cm}{1. Merge the parent tasks' results \\ 2. Compute the new truncated SVD} \\&&& \\
        \hline
        \multirow{2}{1cm}{$Image_E$}&\multirow{2}{1.2cm}{$Task_{20}$}&\multirow{2}{0.45cm}{5}&\multirow{2}{4.2cm}{1. Merge all tasks's results \\ 2. Obtain the final system model} \\&&& \\
		\bottomrule  
	\end{tabular}
\label{Relationships of Tasks in Workflow and Functions}
\end{table}

This operation is therefore named as merge-and-truncate (MAT) operation rather than a simple merging. When there are serval partitions, the MAT operation can be conducted pairwise using a DAG-based strategy as created in Fig. \ref{SVD}. The above steps are summarized in Algorithm \ref{Distributed truncated SVD algorithm}. The matrix $A_{m\times n}$ is first partitioned column-wise and the SVD of each partition is computed in the function \textsc{ParallelSVDbyCols}. The $U, \Sigma, V$s of all partitions are merged using the function \textsc{DoMergeOfBlocks}, which invokes the function \textsc{BlockMerge} as routine. The function \textsc{DoTruncate} carries out the low-rank approximation for $U, \Sigma, V$ by truncating the low-value data in the matrices. The result of truncated SVD is returned by the function \textsc{ParallelSVDbyCols}.

\subsection{DAG establishment of subspace identification}

Based on the above analyses, the SID workflow is established in this subsection. Fig. \ref{Workflow Structure of Subspace Identification Method} provides the workflow structure of subspace identification with the degree of parallelism of SVD stage is $n/col = 10$. Table \ref{Relationships of Tasks in Workflow and Functions} describes the functions of the tasks in the SID workflow and the relationships of them. There are five kinds of task images except the initial task image $Image_{ini}$ used in this workflow. For instance, the image $Image_{C}$ can be reused ten times in the second level of which the task indexes are $Task_{3}-Task_{12}$ but with different input data. In Fig. \ref{Workflow Structure of Subspace Identification Method}, the function and relationship of each kind of the task image can be described as follows.
\begin{figure*}[!ht]
  \centering
  \includegraphics[width=6.4in, height=2.4in]{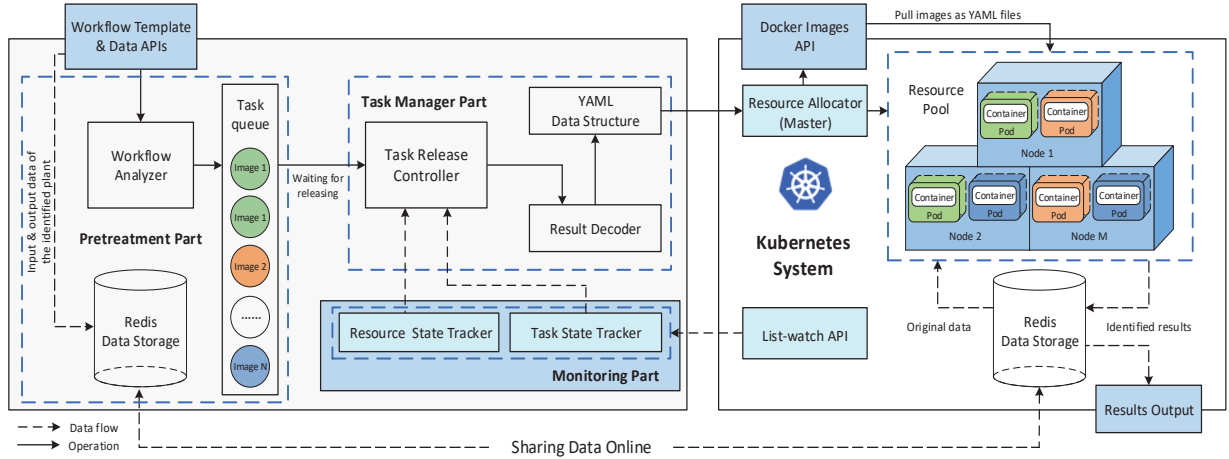}\\
  \caption{The Structure of Containerised Cloud Workflow Processing System}\label{The Structure of Containerised Cloud Workflow Processing System}
\end{figure*}
\begin{enumerate}
  \item The initialization task $Task_{ini}$ is responsible for creating history data of the identified system and preparing the Hankel matrices for the following SID workflow.
  \item Based on the Hankel matrices $[U_p, Y_p, U_f, Y_f]$, $Task_1$ conducts the course of $W_p = [U_p\ Y_p]$, $O_i=f(Y_f,U_f,W_p)$ where the function $f$ means calculating oblique projection as \emph{Definition 2}. The result $O_i$ is divided into ten slices along the column direction.
  \item Similarly, $O_{i-1}$ is calculated by $Task_2$ based on $[U_{p}^{+}, Y_{p}^{+}, U_{f}^{-}, Y_{f}^{-}]$.
  \item The ten slices are delivered to $Task_{3}-Task_{12}$ as input data, respectively. In $Task_{3}-Task_{12}$, the sliced blocks are decomposed by truncated SVD using the function \textsc{DoMergeOfBlocks}.
  \item $Task_{13}-Task_{19}$ carry out the MAT operations on the results of $Task_{3}-Task_{12}$, i.e., merge the parent tasks' results and compute the new truncated SVDs.
  \item Finally, the export task $Task_{20}$ collects all the results of $Task_1$, $Task_2$, $Task_{17}$, $Task_{18}$ and $Task_{19}$ as well as merges them, and later estimates the system model by the least square method.
  \vspace{-0.1cm}
\end{enumerate}

%


In the workflow processing, the MPT is used to represent the maximal parallel tasks in a workflow, which means that at most MPT tasks are running in parallel during the workflow execution. In this case, the MPT is 10 and tasks number is 20. This subsection only provides one case with a specific granularity. Using the same method, other SID workflow structures could be established with different parameters for different kinds and scales of problems and different results would be obtained.

\subsection{Computational complexity analysis}

To study the improvement of this cloud workflow method, the computational complexity of Algorithm \ref{Distributed truncated SVD algorithm} is analyzed. Without loss of generality, we assume all the blocks are divided with the same column width. Let $m\times n$ matrix $A$ is partitioned column-wise into $N$ blocks and the size of each block is $m\times s$ where $s = n/N$. Refer to \cite{bjorck2015numerical}, the floating point operations (flops) number of a complete SVD is approximated as $6mn^{2}+16n^{3}$.

In Algorithm \ref{Distributed truncated SVD algorithm}, the first step is to execute $Image_C$. Thus the flops number of the SVD of each block is $6ms^{2}+16s^{3}$. By the low rank approximation, assume that each SVD is truncated to a $k$-rank matrix. Then at each level of the DAG, the tasks with $Image_D$ are executed. The flops number of $Image_D$ consists of two functions, which are the cost of merging $2ms^{2}$ and the cost of new truncated SVD $4mk^{2}+176k^{3}$, of which the latter is made up by the cost of SVD $6k\cdot k^{2}+16k^{3}=176k^{3}$ and the cost of updating $V$ is $4mk^{2}$.
This task image is required to repeat $N-1$ times. Thus the total flops number is
\begin{eqnarray}\nonumber
  \small{T(m, n, N)=N(6ms^{2}+16s^{3})}&&  \\
    && \!\!\!\!\!\!\!\!\!\!\!\!\!\!\!\!\!\!\!\!\!\!\!\!\!\!\!\!\!\!\!\!\!\!\!\!\!\!\!\!\!\!\!\!\!\!\!\!\! \small{+(N-1)(6mk^{2}+176k^{3})}.
\end{eqnarray}

Since $k\leq s$, the total number of flops
\begin{equation}
  \small{T(m, n, N) < \frac{12mn^{2}}{N}+\frac{192n^{3}}{N^{2}}}.
\end{equation}

Thus, when the degree of parallelism $N$ grows, the proposed method can accelerate the subspace identification effectively. Since most of the tasks can be executed in parallel, the total time cost would be reduced further.

\begin{remark}
In the practical establishment of the SID workflow, we place the last two tasks with $Image_D$ into the export task since the time costs of the two tasks are relatively less. Thus in Fig. \ref{Workflow Structure of Subspace Identification Method}, there are only $7$ tasks with $Image_D$.
\end{remark}

\section{Containerised cloud workflow processing system}\label{Containerised cloud workflow processing system}

This system is the core of cloud platform layer, of which the structure is provided in Fig. \ref{The Structure of Containerised Cloud Workflow Processing System}. Through the entry APIs, the SID workflow template and data are loaded to this system. This system is made up by the pretreatment, task manager and monitoring parts. Then the resource requests are submitted to Kubernetes system and the containers are created and distributed into the nodes in cloud resource pool. In this section, the above three parts and Kubernetes system are described, respectively.

\subsection{Pretreatment part}

This part is in charge of processing the received workflow template as well as data and submitting the results to the subsequent parts. The workflow analyzer parses the workflow template and releases the below information for each task which would be written in YAML files by the task manager.
\begin{itemize}
  \item The task index and the level to this task belongs.
  \item The pre- and post- dependence relationships of this task.
  \item The image which would be pulled from template register.
\end{itemize}

Meanwhile, the input and output data of the identified plant are stored in a redis (remote dictionary server) \cite{nelson2016mastering} data storage system, which could share data online among different hosts via network. The stored data would be loaded to the entrance container when the workflow begins to be executed.

\subsection{Task manager part}

The task manager takes charge of releasing new tasks to cloud resource pool. The task release controller is the kernel module of this part in which the release strategy is imported. This module generates the release command based on the resource usage of the node cluster sensed by the resource state tracker. Then, since Kubernetes system requires the command files in YAML data format, a result decoder is designed to transcode the resource request.

\subsection{Monitoring part}

The monitoring part consists of the resource state tracker and task state tracker, which receive the required states by list-watch scheme \cite{sayfan2017mastering} from Kubernetes system.

\begin{itemize}
  \item The resource state tracker monitors the resource usage of the node cluster to be provided for the task manager.
  \item The task state tracker monitors the states of containers such as running, completed and failed, and informs the task manager if the state of a container changes.
\end{itemize}

\subsection{Kubernetes system}

\begin{figure}[!ht]
  \centering
  \includegraphics[width=3.1in]{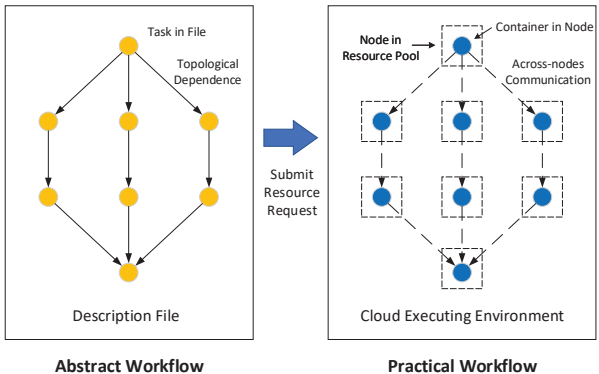}\\
  \caption{Map Relationship Between Abstract and Practical Workflows}\label{The Map Relationship between abstract and practical workflows}
\end{figure}

In this system, the resource allocator or master takes charge of collecting computing resource and creating new containers in node cluster as the requirement of task manager. Then a practical workflow would be created refer to the abstract workflow template as shown in Fig. \ref{The Map Relationship between abstract and practical workflows}. Each task in the abstract workflow would be translated to an executing container scheduled into a node. The inter-dependencies between the tasks would be translated into the across-nodes communications to transmit the intermediate data. In addition, the original data required by the entrance task are provided by the redis data storage system and the identified results are also stored in this redis system which can be accessed by the outside.
\section{Performance Evaluation}\label{Performance Evaluation}

In this section, a simple static algorithm is applied to evaluate the effectiveness of the proposed method and system. Then a series of experiments are carried out and the results as well as discussions are provided.

\subsection{Cloud resource scheduling algorithm}

To keep the load balance of the node cluster, Algorithm \ref{Cloud resource scheduling algorithm for fast subspace identification} is applied in the proposed system. First, the prepared tasks are pushed into the empty $Task\_list$ with their resource requests. Then detect the current load utilization of $N$ computing nodes $\mathcal{U}_1, \mathcal{U}_2, ..., \mathcal{U}_N$. Later, schedule the task to the node with the lowest load utilization and update it. Finally, create the containers in the target nodes as the scheduling solution.

\begin{remark}
In practice, the container can be created and started within one second. However, this process still need cost a little time, which may has influence on the SID missions with strong real-time requirements such as SID-based data driven control. Thus we choose the static algorithm to reduce the influence by creating all containers before the missions start.
\end{remark}

\begin{algorithm}[!ht]
    \renewcommand{\algorithmicrequire}{\textbf{Input:}}
	\renewcommand{\algorithmicensure}{\textbf{Output:}}
    \caption{\textbf{Cloud Resource Scheduling Algorithm For Fast Subspace Identification}}
    \label{Cloud resource scheduling algorithm for fast subspace identification}
    \small{\textbf{Input:} Empty list $Task\_list$ to be scheduled.\\
    \textbf{Output:} Scheduling solution of the $Task\_list$.}

    \begin{algorithmic}[1]
    \small{
         \STATE Put $task_{i}$ with the resource request $r_i$ into $Task\_list$.
         \STATE Detect the load utilization of $N$ nodes $\mathcal{U}_1, \mathcal{U}_2, ..., \mathcal{U}_N$.
         \FOR{$i=1$ to $num(Task\_list)$}
            \STATE Schedule $task_i$ to the node $j$ with the lowest load utilization.
            \STATE Update the load utilization of node $j$, $\mathcal{U}_j = \mathcal{U}_j + r_i$.
            \STATE Remove the task from $Task\_list$.
         \ENDFOR
         \STATE Create the containers in the target nodes.}
    \end{algorithmic}
    \vspace{-0.3em}
\end{algorithm}

\subsection{Experiments setup}

The proposed method is applied to analyze the measurement data from an ball-beam system of which the dynamical model is given as follows:
\begin{equation}
  A \!=\! \left[\!
          \begin{array}{cc}
            2 \!& \!-1 \\
            1 \!& \!0 \\
          \end{array}\!
        \right]\!, B\!=\!\left[\!
                     \begin{array}{c}
                       1 \\
                       0 \\
                     \end{array}\!
                   \right]\!, C \!=\! \left[\!
                                  \begin{array}{cc}
                                    0.00014\! & \!0.00014 \\
                                  \end{array}\!
                                \right]\!, D \!=\! 0
\end{equation}

Three comparative groups including an baseline and two recursive methods and three experiment groups of the proposed cloud-based method are carried out in Alibaba Cloud. The baseline works are executed in the cloud server with 4 CPU and 8 GB memory of which the type is \emph{ecs.hfc6.xlarge}. The recursive method is the main kind of the accelerated SID methods, which improves the computation efficiency by studying numerical algorithms. In this paper, the comparative recursive methods are two state-of-the-art accelerated SID methods, i.e., R4SID \cite{oku2002recursive} and RPBSID$_{\textbf{pm}}$ \cite{houtzager2011recursive}, which increase the computation efficiency by improving the structure of numerical algorithms. The works of the recursive methods are conducted in the same environment with the baseline.

\begin{figure}[!ht]
  \centering
  \includegraphics[width=2.5in]{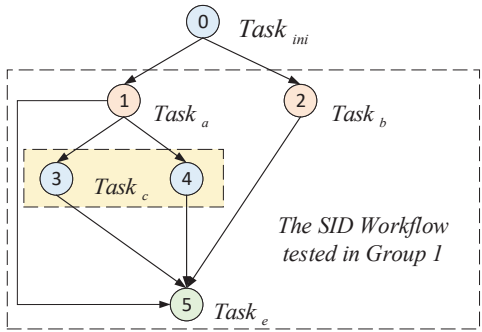}\\
  \caption{The SID workflow Structure Tested in The First Group ($MPT = 2$)}\label{The SID workflow Structure Tested in the First Group}
\end{figure}

The three groups of the cloud-based SID experiments are carried out with different resource configurations or workflow structures. The resources and node cluster information are listed in Table \ref{Resources and node cluster information}. The first and second groups apply two \emph{ecs.hfc6.4xlarge} nodes and each node has 16 CPU and 32 GB memory. Four identical nodes are used in the third group. In addition, two structures of SID workflow are designed to execute the SID missions. The first structure is tested in the first group with the MPT is 2 and tasks number is 5 as shown in Fig. \ref{The SID workflow Structure Tested in the First Group}. The second structure applies the design of Fig. \ref{Workflow Structure of Subspace Identification Method} with the MPT is 10 and tasks number is 20, which is carried out in the second and third groups.

The network in the clusters is set as calico protocol, a scalable network to build the connections between containers. The version of redis is 3.5.3. The containerised cloud workflow processing system is implemented by Golang of which the version is 1.10.4. The tasks of the SID workflows are implemented by Python of which the version is 3.7. The basic scientific libraries used in the implementation of the SID workflows are Numpy (1.19.5) and Scipy (1.5.4), respectively. The operation system is Ubuntu 20.04. All the experiments are repeated as Monte Carlo method of which the averaged results are recorded in Table \ref{Recorded results statistics for topical parameters in Monte Carlo experiments}.

\begin{table*}[htbp]
	\centering
	\caption{Recorded results statistics for topical parameters in Monte Carlo experiments}
    \renewcommand{\multirowsetup}{\centering}
	\begin{tabular}{|c|c|c|c|c|c|c|c|c|}
		\hline  
        \cline{1-9}
        \multicolumn{3}{|c|}{\multirow{4}{*}{\textbf{Parameters}}}& \multicolumn{2}{c|}{\multirow{2}{*}{\textbf{Middle-scale}}} & \multicolumn{4}{c|}{\multirow{2}{*}{\textbf{Large-scale}}} \cr
        \multicolumn{3}{|c|}{\multirow{4}{*}{}}& \multicolumn{2}{c|}{\multirow{2}{*}{}} & \multicolumn{4}{c|}{\multirow{2}{*}{}} \cr

        \cline{4-9}
        \multicolumn{3}{|c|}{\multirow{2}{*}{}}&\multirow{2}{1.55cm}{$N=10$\\$j=1000$}&\multirow{2}{1.55cm}{$N=20$\\$j=1000$}&\multirow{2}{1.55cm}{$N=10$\\$j=10000$}&\multirow{2}{1.55cm}{$N=20$\\$j=10000$}&\multirow{2}{1.55cm}{$N=50$\\$j=10000$}&\multirow{2}{1.55cm}{$N=50$\\$j=20000$}\cr
        \multicolumn{3}{|c|}{\multirow{2}{*}{}}&\multirow{2}{*}{}&\multirow{1}{1.5cm}{}&\multirow{2}{1.5cm}{}&\multirow{2}{1.5cm}{}&\multirow{2}{1.5cm}{}&\multirow{2}{1.5cm}{}\cr
        \hline

        \hline\hline
        \multicolumn{2}{|c|}{\multirow{1}{*}{\textbf{Baseline}}}&\multirow{1}{1.55cm}{Time (s)}&
        \multirow{1}{*}{0.0151}&\multirow{1}{*}{0.0245}&\multirow{1}{*}{1.8402}&\multirow{1}{*}{2.4803}&\multirow{1}{*}{4.4312}&\multirow{1}{*}{18.1912} \cr

        \hline
        \multicolumn{2}{|c|}{\multirow{2}{2.6cm}{\textbf{Recursive Method 1: \\R4SID}}}&\multirow{1}{*}{Time (s)}&
        \multirow{1}{*}{0.1253}&\multirow{1}{*}{0.1765}&\multirow{1}{*}{1.1914}&\multirow{1}{*}{1.4675}&\multirow{1}{*}{5.8254}&\multirow{1}{*}{11.3162} \cr

        \cline{3-9}
        \multicolumn{2}{|c|}{\multirow{2}{*}{\textbf{}}}&\multirow{1}{*}{Reduction}
        &\multirow{1}{*}{-729.8\%}&\multirow{1}{*}{-583.7\%}&\multirow{1}{*}{35.3\%}&\multirow{1}{*}{40.8\%}&\multirow{1}{*}{-31.5\%}&\multirow{1}{*}{41.0\%} \cr

        \hline
        \multicolumn{2}{|c|}{\multirow{2}{2.6cm}{\textbf{Recursive Method 2: \\ RPBSID$_{\textbf{pm}}$}}}&\multirow{1}{*}{Time (s)}&
        \multirow{1}{*}{0.1102}&\multirow{1}{*}{0.1278}&\multirow{1}{*}{1.0965}&\multirow{1}{*}{1.2208}&\multirow{1}{*}{4.0482}&\multirow{1}{*}{7.9165} \cr

        \cline{3-9}
        \multicolumn{2}{|c|}{\multirow{2}{*}{\textbf{}}}&\multirow{1}{*}{Reduction}
        &\multirow{1}{*}{-629.8\%}&\multirow{1}{*}{-421.6\%}&\multirow{1}{*}{40.4\%}&\multirow{1}{*}{50.8\%}&\multirow{1}{*}{8.6\%}&\multirow{1}{*}{56.5\%} \cr

        \hline\cline{1-9}
        \multicolumn{1}{|c|}{\multirow{4}{2.35cm}{\textbf{Group 1: \\MPT = 2\\ Tasks = 5}}}&\multirow{2}{1.2cm}{4CPU\\8GB}&\multirow{1}{*}{Time (s)}&
        \multirow{1}{*}{\textbf{0.0114}}&\multirow{1}{*}{\textbf{0.0164}}&\multirow{1}{*}{0.4867}&\multirow{1}{*}{0.7527}&\multirow{1}{*}{3.1082}&\multirow{1}{*}{13.8380} \cr

        \cline{3-9}
        \multicolumn{1}{|c|}{\multirow{4}{1.5cm}{\textbf{}}}&\multirow{2}{*}{}&\multirow{1}{*}{Reduction}&
        \multirow{1}{*}{24.5\%}&\multirow{1}{*}{33.1\%}&\multirow{1}{*}{73.6\%}&\multirow{1}{*}{69.7\%}&\multirow{1}{*}{29.9\%}&\multirow{1}{*}{23.9\%} \cr

        \cline{2-9}
        \multicolumn{1}{|c|}{\multirow{4}{1.5cm}{\textbf{}}}&\multirow{2}{0.8cm}{8CPU\\8GB}&\multirow{1}{*}{Time (s)}&
        \multirow{1}{*}{\textbf{0.0112}}&\multirow{1}{*}{\textbf{0.0163}}&\multirow{1}{*}{0.4369}&\multirow{1}{*}{0.6543}&\multirow{1}{*}{3.0923}&\multirow{1}{*}{12.8730} \cr

        \cline{3-9}
        \multicolumn{1}{|c|}{\multirow{4}{1.5cm}{\textbf{}}}&\multirow{2}{*}{}&\multirow{1}{*}{Reduction}&
        \multirow{1}{*}{25.8\%}&\multirow{1}{*}{33.5\%}&\multirow{1}{*}{76.3\%}&\multirow{1}{*}{73.6\%}&\multirow{1}{*}{30.2\%}&\multirow{1}{*}{29.2\%} \cr

        \hline\cline{1-9}
        \multicolumn{1}{|c|}{\multirow{4}{1.5cm}{\textbf{Group 2: \\MPT = 5\\ Tasks = 20}}}&\multirow{2}{0.8cm}{4CPU\\8GB}&\multirow{1}{*}{Time (s)}&
        \multirow{1}{*}{0.0938}&\multirow{1}{*}{0.1424}&\multirow{1}{*}{0.3692}&\multirow{1}{*}{0.6859}&\multirow{1}{*}{1.3578}&\multirow{1}{*}{2.7693} \cr

        \cline{3-9}
        \multicolumn{1}{|c|}{\multirow{4}{1.5cm}{\textbf{}}}&\multirow{2}{*}{}&\multirow{1}{*}{Reduction}&
        \multirow{1}{*}{-521.2\%}&\multirow{1}{*}{-481.2\%}&\multirow{1}{*}{79.9\%}&\multirow{1}{*}{72.3\%}&\multirow{1}{*}{69.4\%}&\multirow{1}{*}{84.8\%} \cr

        \cline{2-9}
        \multicolumn{1}{|c|}{\multirow{4}{1.5cm}{\textbf{}}}&\multirow{2}{0.8cm}{8CPU\\8GB}&\multirow{1}{*}{Time (s)}&
        \multirow{1}{*}{0.1019}&\multirow{1}{*}{0.1233}&\multirow{1}{*}{0.3517}&\multirow{1}{*}{0.6776}&\multirow{1}{*}{1.3331}&\multirow{1}{*}{2.7892} \cr

        \cline{3-9}
        \multicolumn{1}{|c|}{\multirow{4}{1.5cm}{\textbf{}}}&\multirow{2}{*}{}&\multirow{1}{*}{Reduction}&
        \multirow{1}{*}{-574.8\%}&\multirow{1}{*}{-403.3\%}&\multirow{1}{*}{80.9\%}&\multirow{1}{*}{72.7\%}&\multirow{1}{*}{69.9\%}&\multirow{1}{*}{84.7\%} \cr

        \hline\cline{1-9}
        \multicolumn{1}{|c|}{\multirow{4}{1.5cm}{\textbf{Group 3: \\MPT = 5\\ Tasks = 20}}}&\multirow{2}{0.8cm}{4CPU\\8GB}&\multirow{1}{*}{Time (s)}&
        \multirow{1}{*}{0.0553}&\multirow{1}{*}{0.0663}&\multirow{1}{*}{0.1564}&\multirow{1}{*}{0.5110}&\multirow{1}{*}{0.7554}&\multirow{1}{*}{1.7625} \cr

        \cline{3-9}
        \multicolumn{1}{|c|}{\multirow{4}{1.5cm}{\textbf{}}}&\multirow{2}{*}{}&\multirow{1}{*}{Reduction}&
        \multirow{1}{*}{-266.2\%}&\multirow{1}{*}{-170.6\%}&\multirow{1}{*}{\textbf{91.5\%}}&\multirow{1}{*}{\textbf{79.4\%}}&\multirow{1}{*}{\textbf{83.0\%}}&\multirow{1}{*}{\textbf{90.3\%}} \cr

        \cline{2-9}
        \multicolumn{1}{|c|}{\multirow{4}{1.5cm}{\textbf{}}}&\multirow{2}{0.8cm}{8CPU\\8GB}&\multirow{1}{*}{Time (s)}&
        \multirow{1}{*}{0.0823}&\multirow{1}{*}{0.0747}&\multirow{1}{*}{0.1552}&\multirow{1}{*}{0.5161}&\multirow{1}{*}{0.7926}&\multirow{1}{*}{1.8851} \cr

        \cline{3-9}
        \multicolumn{1}{|c|}{\multirow{4}{1.5cm}{\textbf{}}}&\multirow{2}{*}{}&\multirow{1}{*}{Reduction}&
        \multirow{1}{*}{-445.0\%}&\multirow{1}{*}{-204.9\%}&\multirow{1}{*}{\textbf{91.6\%}}&\multirow{1}{*}{\textbf{79.2\%}}&\multirow{1}{*}{\textbf{82.1\%}}&\multirow{1}{*}{\textbf{89.6\%}} \cr

        \hline
        \cline{1-9}
	\end{tabular}
\label{Recorded results statistics for topical parameters in Monte Carlo experiments}
\end{table*}

\begin{table}[htbp]
	\centering
	\caption{Resources and node cluster information}
	\begin{tabular}{|c|c|c|c|}
		\toprule  
		\textbf{Experiment Group}&\textbf{Nodes}&\textbf{Total Resource Pool}&\textbf{Node Type}  \\
		\hline  
        \hline
        Baseline&1 & 4 CPU, 8GB&\emph{ecs.hfc6.xlarge} \\
        \hline
        Recursive&1 & 4 CPU, 8GB&\emph{ecs.hfc6.xlarge} \\
        \hline
        Group 1&2 & 32 CPU, 64GB&\emph{ecs.hfc6.4xlarge}\\
        \hline
        Group 2&2 & 32 CPU, 64GB& \emph{ecs.hfc6.4xlarge}\\
        \hline
        Group 3&4 & 64 CPU, 128GB &\emph{ecs.hfc6.4xlarge}\\
		\bottomrule  
	\end{tabular}
\label{Resources and node cluster information}
\end{table}

\subsection{Results and discussion}

The experiment results including computation times and reduction percentage are presented in Table \ref{Recorded results statistics for topical parameters in Monte Carlo experiments}. Furthermore, the related discussion about the acceleration efficiency and three factors consisting of data scale, resource amount and container configuration, and the comparison with the recursive methods are provided as the follows.

\vspace{0.1cm}\noindent\emph{C1. Acceleration efficiency}\vspace{0.1cm}

From Table \ref{Recorded results statistics for topical parameters in Monte Carlo experiments}, the experiment results can be divided into two parts as parameters range: the middle- (first two rows) and large- (last four rows) scales. For the two kinds of parameters, different acceleration efficiencies are achieved.

\begin{figure}[htbp!]
  \centering
  \includegraphics[width=3.6in, height=1.55in,trim=20 6 56 25,clip]{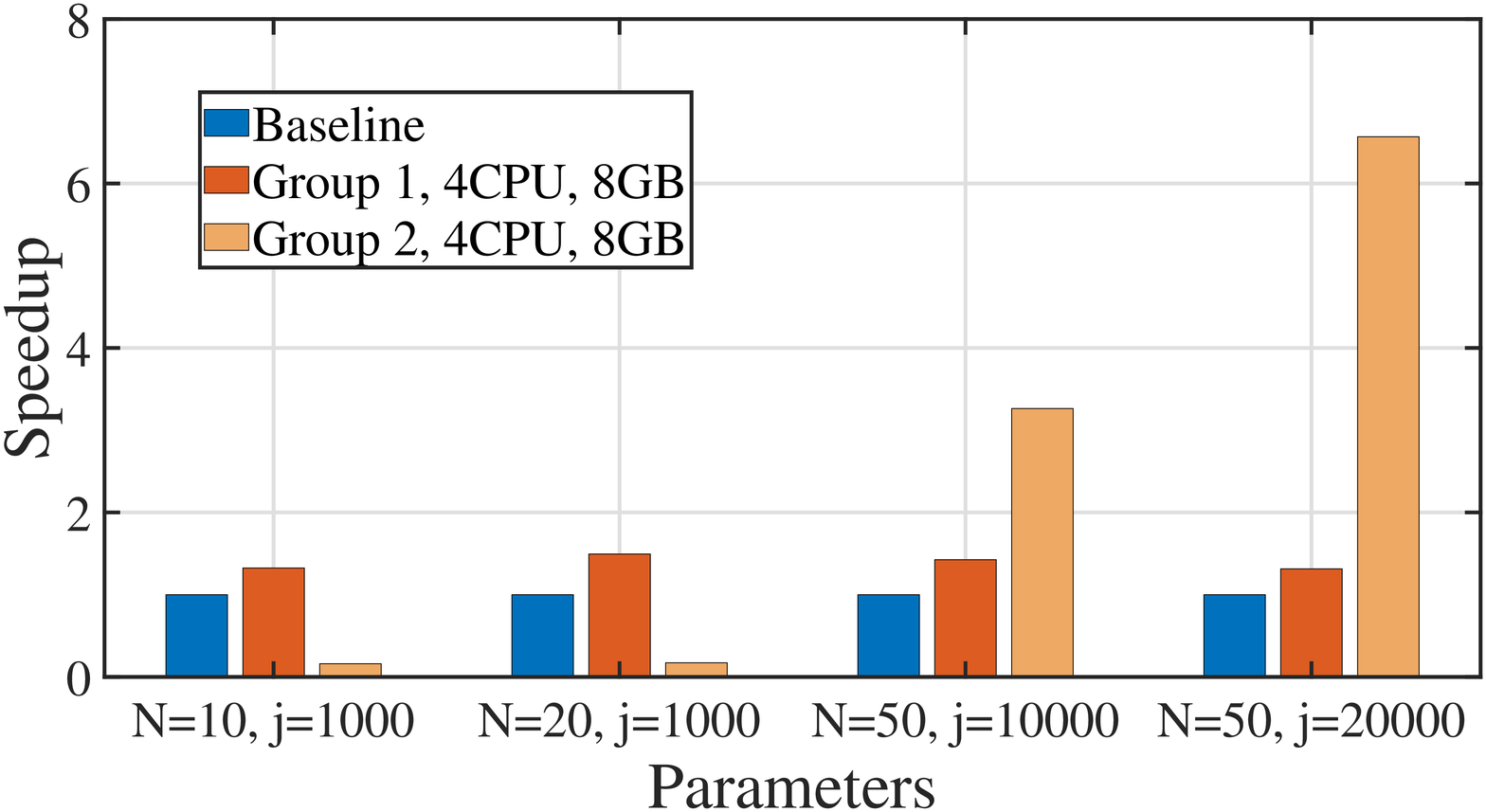}\\
  \vspace{-0.2cm}
  \caption{The Relationship Between Speedup and Workflow Granularity}\label{Relationship Between Speedup and Workflow Granularity}
\end{figure}

\begin{figure}[htbp!]
  \centering
  \includegraphics[width=3.6in, height=1.55in,trim=20 6 56 25,clip]{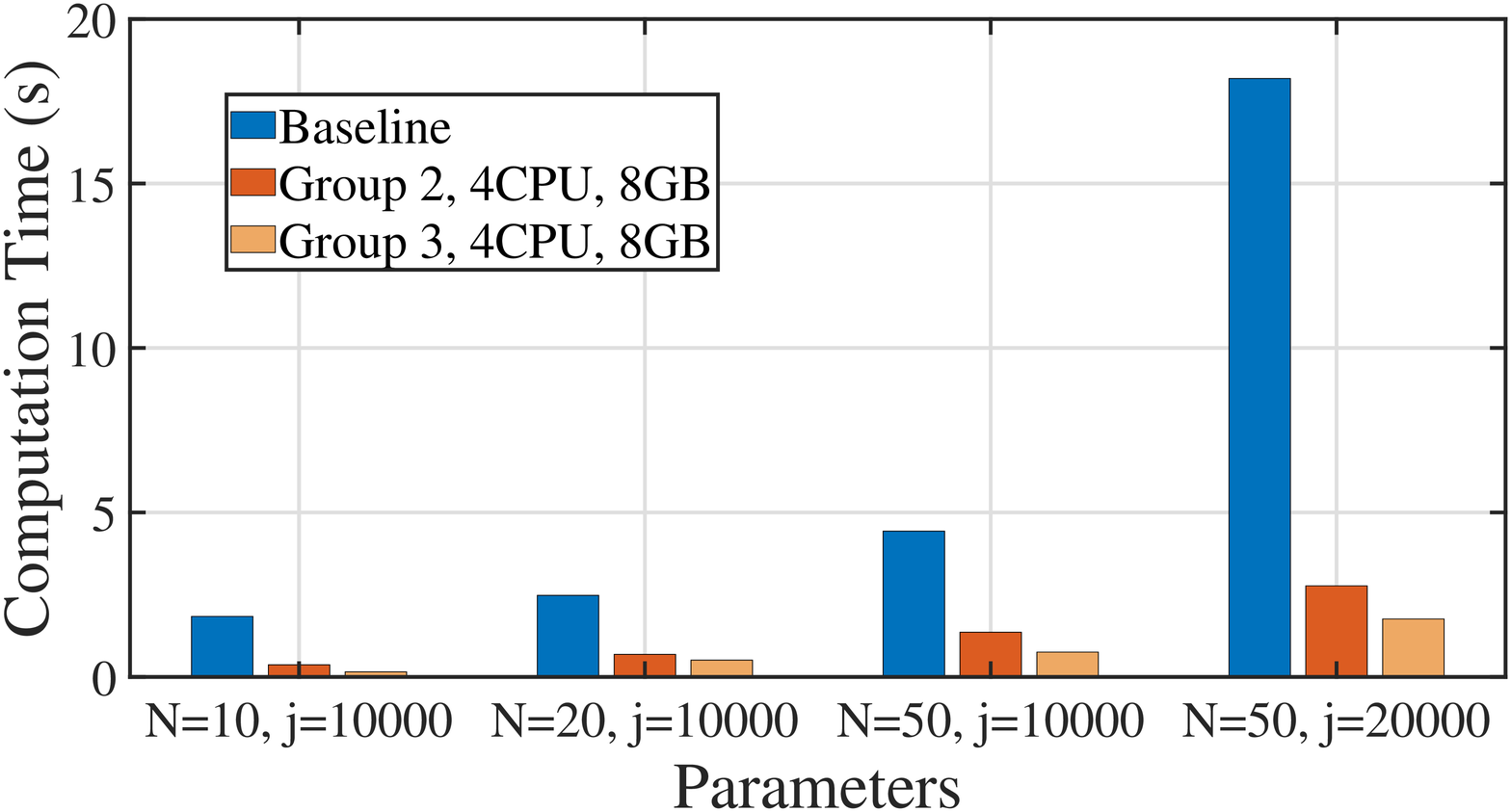}\\
  \vspace{-0.2cm}
  \caption{The Relationship Between Computation Time and Resource Amount}\label{Relationship Between Computation time and Total Resource Amount}
\end{figure}

\begin{figure}[htbp!]
  \centering
  \includegraphics[width=3.6in, height=1.55in,trim=20 6 56 25,clip]{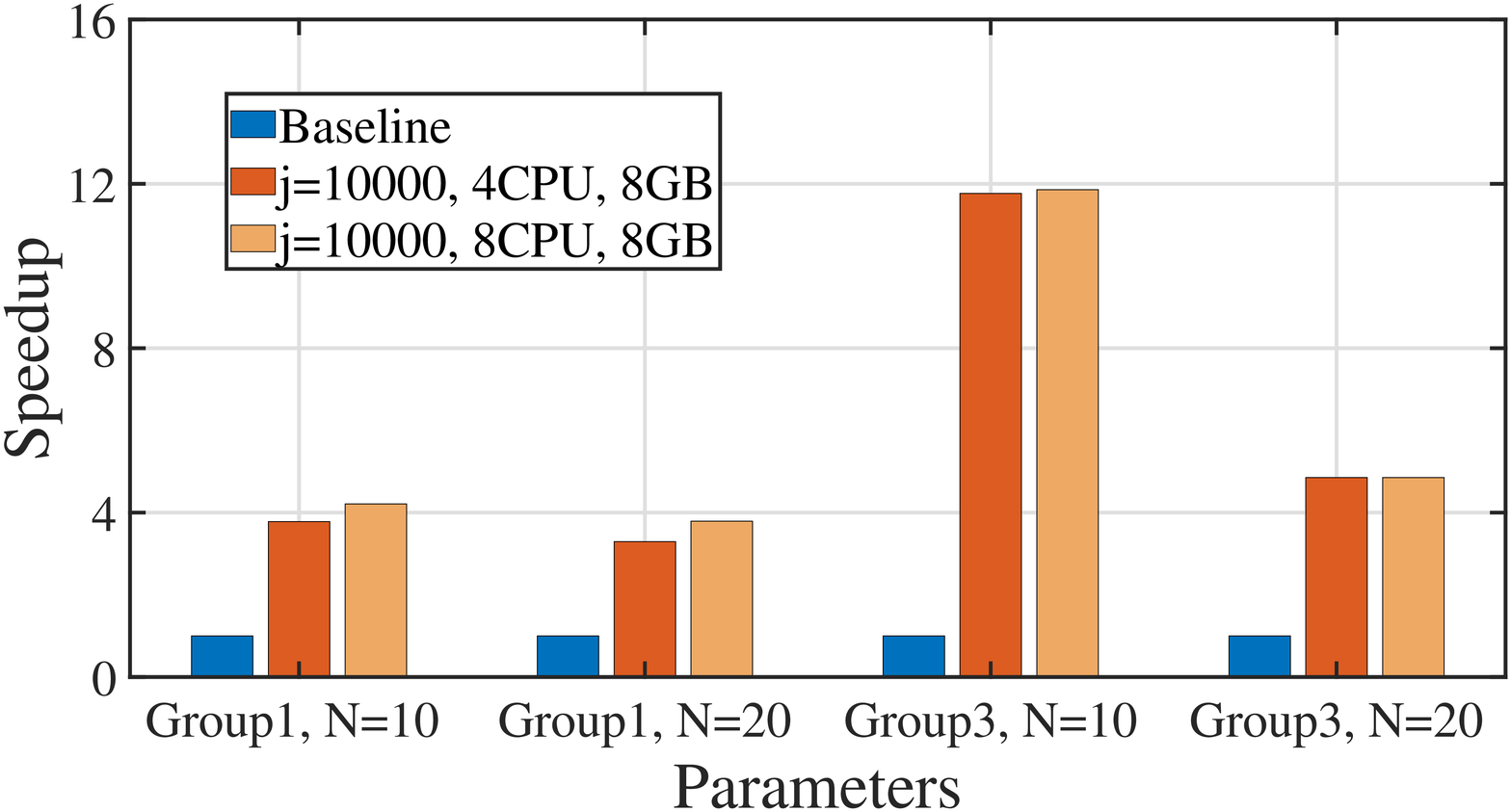}\\
  \vspace{-0.2cm}
  \caption{The Relationship Between Speedup and Container Configuration}\label{Relationship Between Speedup and Container Configuration}
\end{figure}

\begin{enumerate}
  \item For the middle-scale parameters, the computation times are reduced to within 20 ms which is usually the period of real-time dynamical control \cite{nilsson1998real}. That is to say, the proposed fast SID method can be applied to the data-driven control with high-frequency requirements such as vehicle control with no need of delay compensation.
  \item For the large-scale parameters, the computation times are reduced by at most $91.6\%$, which would be very useful for the identification with high-volume data and high-complexity such as power grid and multi-agents.
\end{enumerate}

\vspace{0.12cm}\noindent\emph{C2. Granularity of workflow strcutre}\vspace{0.12cm}

In parallel system, the speedup is defined as the ratio of the data processing time on a single processor and that on multi-processors expressed as
\begin{equation}
  S_p = \frac{T_{s}}{T_{p}}
\end{equation}
where $T_s$ is the processing time of sequential execution and $T_p$ is that of parallel execution. The relationship between the speedup and workflow granularity is presented in Fig. \ref{Relationship Between Speedup and Workflow Granularity}. For the middle-scale parameters, the first workflow structure with the MPT = 2 achieved better performance while the second structure with the MPT = 10 significantly decreases the computation speed. This is because the data amount processed in this scale is relatively small and more time are costed in communication when applying the second structure with much more channels. For the large-scale parameters, the data amount processed in each container is larger and thus the high-MPT structure brought significant improvement. This case states that the computation efficiency could be improved if applying the workflow structure with proper granularity.

\vspace{0.18cm}\noindent\emph{C3. Resource amount}\vspace{0.12cm}

The relationship between the computation time and total resource amount is shown in Fig. \ref{Relationship Between Computation time and Total Resource Amount}. It's clear that all the computation times are reduced with the total resource amount rising. For example, the baseline is 18.1912 s when $N=50$, $j=20000$. The computation time is reduced to 2.7692 s and the reduction percentage is $84.8\%$ when the cloud resource pool had 32 CPU and 64 GB memory. The computation time is further reduced to 1.7625 s and the reduction percentage is $89.6\%$ when the total resources are 64 CPU and 128 GB memory. This case states that a high computation speed could be achieved if enough computing resources are provided.

\vspace{0.18cm}\noindent\emph{C4. Container configuration}\vspace{0.12cm}

The comparison experiments are conducted with different container configurations for the above three groups. Since the SVD is the main time-consuming stage, the containers processing the SVD tasks are set as 8 CPU and 8 GB memory. However, the speedup improvement is limited as shown in Fig. \ref{Relationship Between Speedup and Container Configuration}. In the first two bar charts, the speedups improved from 3.7810 to 4.2119 and from 3.2952 to 3.7908, respectively. Since there are only 2 SVD tasks in the first group, the resources are relatively unstrained. Therefore the upgrade of container configuration could lead some performance improvement.

But in the latter two bar charts, the speedups remained nearly unchanged. This is because there are 17 SVD tasks in the third group and the resources become strained. When the containers are created and run in cloud resource pool at the same time, the competition for resources occurs and the containers affect each other. This case states that the improvement of upgrading container configuration is depended on whether the current resources are sufficient.

\vspace{0.18cm}\noindent\emph{C5. Comparison with the recursive methods}\vspace{0.12cm}

The recursive SID methods are the typical accelerated SID methods based on the improvement of the numerical algorithm design. Thus the recursive methods are used to compare the computation efficiency of the traditional numerical improvement and cloud-based methods. Provided the parameters of middle- and large- scales, the recursive methods are also conducted in the same environment with the baseline. The results of Monte Carlo experiments are recorded in Table. \ref{Recorded results statistics for topical parameters in Monte Carlo experiments}.

From this Table, the results are divided into middle-scale and large-scale parts. In the middle-scale parameters, the recursive methods decrease the computation speed of the SID missions, of which the computation times are about several times that of the baseline. This is because the data scale is relatively small and the iteration numbers of the recursive method are high, which lead the waste of computing resource. As the comparison, the cloud-based SID method with the workflow structure of Fig. \ref{The SID workflow Structure Tested in the First Group} obtains better results which can satisfy the real-time requirements. This case states that the cloud-based method could achieve faster computation speed for real-time mission by designing the workflow structure with proper granularity.

In the large-scale parameters, the recursive methods reduce the computation time of the SID mission in most cases. The reduction percentages are from $35\%$ to $41\%$ and $40\%$ to $57\%$, respectively. But the computation speed is sensitive to the length-width ratio of the matrix. For the R4SID, when $N=50, j=10000$, i.e., the length-width ratio becomes larger, the computation time increases $31.5\%$. For the RPBSID$_{\textbf{pm}}$, the  reduction percentage of the computation time is only $8.6\%$. As the comparison, the computation times are reduced by at most $91.6\%$ by the cloud-based SID method. This case states that the proposed method could significantly improve the upper bound of the computation speed by cloud computing.



\section{Conclusion}\label{Conclusion}
This paper has presented a cloud-based method of fast SID combing workflow processing approach and container technology, which could reduce the computation time by at most $91.6\%$ for large-scale SID missions and satisfy the real-time requirement for the parameter of SID-based data driven predictive control. The proposed method can well deal with the large-scale and real-time identification missions which are the main difficulties of the current SID methods. In the future, cloud-based fast data driven control and model predictive control would be studied based on this work.

%

\begin{appendices}
\section{Definitions in N4SID Algorithm}\label{Definitions in N4SID Algorithm}
In what follows, the matrices $\mathcal{A}\in \mathbb{R}^{p \times j}$ and $\mathcal{B}\in \mathbb{R}^{q \times j}$.

\emph{\textbf{Definition 1:} Orthogonal projection.}

The orthogonal projection of the row space of $\mathcal{A}$ into the row space of $\mathcal{B}$ is denoted by $\mathcal{A}/\mathcal{B}$ and defined as:\vspace{-0.1cm}
\begin{equation}
  \small{\mathcal{A}/\mathcal{B} = \mathcal{A}\mathcal{B}^{\dagger}\mathcal{B}}
  \vspace{-0.15cm}
\end{equation}
where $\dag$ means the Morre-Penrose pesudo-inverse.

$\mathcal{A}/\mathcal{B}^{\bot}$ is the projection of the row space of $\mathcal{A}$ into $\mathcal{B}^{\bot}$ where $\mathcal{B}^{\bot}$ represents the orthogonal complement of the row space of $\mathcal{B}$, for which we have $\mathcal{A}/\mathcal{B}^{\bot}=\mathcal{A}-\mathcal{A}/\mathcal{B}$.

\emph{\textbf{Definition 2:} Oblique projection.}

The oblique projection of the row space of $\mathcal{A}$ along the row space of $\mathcal{B}$ into the row space $\mathcal{C}\in \mathbb{R}^{r \times j}$ is defined as:\vspace{-0.1cm}
\begin{equation}
  \small{\mathcal{A}/_{\mathcal{B}}\,\mathcal{C}=\left(\!
                                            \begin{array}{c}
                                              \mathcal{A}/\mathcal{B}^{\bot} \\
                                            \end{array}
                                          \!\right)\left(\!
                                            \begin{array}{c}
                                              \mathcal{C}/\mathcal{B}^{\bot} \\
                                            \end{array}
                                          \!\right)^{\dagger}\mathcal{C}}.
                                          \vspace{-0.1cm}
\end{equation}

$U_{p}^{+}$ and $U_{f}^{-}$ are defined as\vspace{-0.1cm}
\begin{eqnarray}
\centering
  \small{\!\!\!\!\!\!U_{p}^{+} \!\!\!\!}&=& \small{\!\!\!\!\!\!\left[
            \begin{array}{cccc}
              \!\!u(0) & \!\!u(1) & \!\!\ldots & \!\!u(j-1) \\
              \!\!u(1) & \!\!u(2) & \!\!\ldots & \!\!u(j) \\
              \!\!\vdots & \!\!\vdots & \!\!\ddots & \!\!\vdots \\
              \!\!u(N) & \!\!u(N+1) & \!\!\ldots & \!\!u(N+j-1) \\
            \end{array}
          \right]}, \\
  \small{\!\!U_{f}^{-} \!\!\!\!}&=& \small{\!\!\!\!\!\!\left[
            \begin{array}{cccc}
              \!\!\!u(N+1) & \!\!\!u(N+2) & \!\!\!\ldots & \!\!\!u(N+j) \\
              \!\!\!u(N+2) & \!\!\!u(N+3) & \!\!\!\ldots & \!\!\!u(N+j+1) \\
              \!\!\!\vdots & \!\!\!\vdots & \!\!\!\ddots & \!\!\!\vdots \\
              \!\!\!u(2N-1) & \!\!\!u(2N) & \!\!\!\ldots & \!\!\!u(2N+j-2) \\
            \end{array}
          \right]}
          \vspace{-0.2cm}
\end{eqnarray}
where $U_{p}^{+}$ has one more vector row than $U_{p}$ and $U_{f}^{-}$ has one less vector row than $U_{f}$. $Y_{p}^{+}$ and $Y_{f}^{-}$ are denoted similarly.

Then, $\textbf{W}_p^{+}$ is defined as
\begin{equation}
  \small{\textbf{W}_p^{+} = \left[
          \begin{array}{c}
            Y_{p}^{+} \\
            U_{p}^{+} \\
          \end{array}
        \right]}.
\end{equation}

In the step 5, $Y_i$ and $U_i$ are defined as
\begin{eqnarray}
  \small{Y_i \!} &=& \small{\!\!\left[
                       \begin{array}{cccc}
                         y(N) & y(N\!+\!1) & \cdots & y(N\!+\!j\!-\!1) \\
                       \end{array}
                     \right]}
   \\
  \small{U_i \!} &=& \small{\!\!\left[
                       \begin{array}{cccc}
                         u(N) & u(N\!+\!1) & \cdots & u(N\!+\!j\!-\!1) \\
                       \end{array}
                     \right]}
\end{eqnarray}


\section{Step 4 in N4SID Algorithm}\label{Step 4 in N4SID Algorithm}

The extended matrices $\Gamma^{\dag}_{i}$ and $\Gamma^{\dag}_{i-1}$ and estimated state sequences $X_{i}$ and $X_{i+1}$ are determined as $\small{\Gamma^{\dag}_{i}= W_{1}^{-1}U_{1}S_{1}^{1/2}, \Gamma^{\dag}_{i-1} = \underline{\Gamma^{\dag}_{i}}, X_{i} = \Gamma^{\dag}_{i}\mathcal{O}_{i}, X_{i+1} = \Gamma^{\dag}_{i-1}\mathcal{O}_{i-1}}$ where $\underline{\Gamma_{i}}$ means $\Gamma_{i}$ without the last block row.


\end{appendices}

\bibliographystyle{ieeetr}
\bibliography{IEEEabrv,mybib}

\end{document}